\magnification=\magstep1
\baselineskip=16pt
\hfuzz=6pt
$ $

\vskip 1in

\centerline{\bf  The quantum Goldilocks effect:}

\centerline{\it on the convergence of timescales in quantum
transport}

\vskip 1cm

\centerline{Seth Lloyd$^{1,2}$, Masoud Mohseni$^2$, Alireza Shabani$^3$,
Herschel Rabitz$^3$}

\bigskip

\centerline{1. Department of Mechanical Engineering, 2. Research Lab
for Electronics}

\centerline{Massachusetts Institute of Technology}

\centerline{3. Department of Chemistry, Princeton University}

\vskip 1cm

\noindent{\it Abstract:} Excitonic transport in photosynthesis
exhibits a wide range of time scales.  Absorption and initial
relaxation takes place over tens of femtoseconds.  Excitonic
lifetimes are on the order of a nanosecond.  Hopping rates, energy
differences between chromophores, reorganization energies, and
decoherence rates correspond to time scales on the order of
picoseconds. The functional nature of the divergence of time scales
is easily understood: strong coupling to the electromagnetic field
over a broad band of frequencies yields rapid absorption, while long
excitonic lifetimes increase the amount of energy that makes its way
to the reaction center to be converted to chemical energy.  The
convergence of the remaining time scales to the centerpoint of the
overall temporal range is harder to understand.  In this paper we
argue that the convergence of timescales in photosynthesis can be
understood as an example of the `quantum Goldilocks effect': natural
selection tends to drive quantum systems to the degree of quantum
coherence that is `just right' for attaining maximum 
efficiency. We provide a general theory of optimal and robust, efficient 
transport in quantum systems, and show that it is governed by a single
parameter.

\vfill\eject

Photosynthetic complexes are, by definition, complex [1].
They contain
many light harvesting sub-complexes, each of which in turn contains
multiple chromophores for storing and transporting excitons.
Associated with the hierarchical arrangement of photosynthetic
complexes is a diverse spread of time scales.  The energy of
the photons to be absorbed, and of the excitons created
in the complex, corresponds to time scales of femtoseconds.
The relaxation of the excitons to their ground states within
a chromophore takes place over tens of femtoseconds.  As noted
above, a large number of processes in
photosynthetic complexes take place over a fraction of a picosecond to
several picoseconds, including hopping between neighboring
chromophoric complexes, reorganization and relaxation
in the bath of phonons, and decoherence.  Similarly,
energy differences between excitonic states of chromophores
in different molecular environments typically correspond
to picosecond time scales.  The overall transfer
time from absorption in the antenna to capture in the reaction
center is tens of picoseconds.  The entire process
is limited by the excitonic lifetime, which is on the order
of a nanosecond.

The time scales involved in photosynthetic processes span six orders
of magnitude, from femtoseconds to nanoseconds, with a convergence
of the time scales for many processes in the center of this range.
Note that the these convergent processes are exactly those where the
interplay between quantum coherence and decoherence in multi-body
interactions can play an important role. Recent experiments,
followed by detailed theoretical analyses, show that this interplay
is crucial to excitonic transport in photosynthesis [2-10]. The
purpose of this paper is to explain both the divergence of time
scales in photosynthesis, and the convergence of the time scales for
the particularly quantum processes at the picosecond range.

The functional purpose of divergent time scales is easy to understand.
Efficient absorption of light over a broad band of frequencies
requires both strong coupling of light to excitons, i.e.,
large induced dipole moments, and broad bandwidths for
absorption, i.e., rapid relaxation.  The stronger the coupling,
and the broader the absorptive bandwidth, the more energy
available to the bacterium or plant performing photosynthesis.
Accordingly, we should not be surprised that over time, natural
selection leads to fast time scales associated with absorption
of light and creation of excitons.  Similarly, the large separation
between the time scale for creation of excitons and the time scale
for their eventual decay should come as no surprise: the longer
the excitonic lifetime, the more likely it is to make it from
the photosynthetic antenna to the reaction center before decaying.

The functional purpose of the convergence of time scales is
harder to understand, for the simple reason that convergence
of time scales makes quantum systems hard to model.  Our ability
to make simplified, perturbative models of complex quantum systems
hinges crucially on the separation of time and energy scales,
so that quantum effects with divergent time scales can be regarded
as perturbations on each other.  From the perspective
of a scientist trying to analyze excitonic transport in
photocomplexes, the convergence of time scales seems almost
to be an effort on the part of nature to frustrate our
understanding.  Nature is modest and keeps her secrets well.
Independent of nature's intrinsic modesty, however, the
convergence of time scales can play a functional role in
enhancing excitonic transport -- or any other form of
transport.  When the time scales for two processes converge --
e.g., coherent tunneling time and decoherence time --
then the two processes affect each other strongly.
The convergence of time scales can then either assist
energy transport, or interfere with it.  In naturally
occurring systems that have undergone a long process
of dynamic refinement via natural selection, the convergent
processes typically help each other out.

Recently, we developed a non-perturbative, non-Markovian master
equation technique for simulating the behavior of complex quantum
systems over a wide variety of time and energy scales [9].  We
applied this technique to the Fenna-Matthews-Olson complex (FMO), a
seven-chromophore energy transport complex in green sulphur bacteria
[10].  In that work, we found that the convergence of timescales in
FMO is tuned to give high transport efficiency (virtually 100 \%)
that is robust over several decades of variation in the underlying
parameters of the system. A similar study has also shown the
optimality of energy transfer efficiency in the FMO complex [11].
Moreover, the efficiency of transport was shown in [9-10] to be a
function of a single underlying parameter that is linear in the
various timescales themselves. The purpose of this paper is give a
heuristic but quantitative derivation for how convergence of
timescales yields optimal and robust transport.

The paper is organized as follows.  First, we review the classical
Goldilocks principle for complex, designed systems, explaining why
it is important to attain just the right level of complexity
[12-14].  Second, we review the theory of environmentally assisted
quantum transport (ENAQT), which gives the underlying theoretical
mechanism for robust and efficient excitonic transport [5-8].
Finally, we give a heuristic analysis of how different timescales
enter into the transport process to derive the quantum Goldilocks
principle: efficiency is optimized at just the right level
of quantum coherence.  We confirm the existence of a
single parameter that governs transport in complex, partially
ordered quantum systems [9-10], and show why this parameter takes
its particular form as a ratio of products of the different
time and energy scales in the system.  Because of the simple
functional form of the parameter governing the efficiency of transport,
optimal transport rates -- approaching $100\%$ in biological
systems -- can be readily attained by natural selection.
At any point of evolution, there are many paths towards
the global optimum of transport efficiency, and 
there are no local optima in which to be trapped along the way.

\bigskip\noindent{\it The Goldilocks principle for complex,
designed systems}

The story of Goldilocks and the three bears was published
by the English poet Robert Southey in 1837, although it
is presumably a variation on a much older folktale.
In the version known today, a small girl with blonde
hair, Goldilocks, is lost in a forest.  She finds a house
whose inhabitants have just left it.  The house contains
three of everything -- there are three chairs, three
bowls of porridge sitting on a table, three beds, etc.
One of the chairs is too big for Goldilocks, one too small,
and one just the right size.  So she sits on the one that is just
right.  One of the bowls of porridge is too hot, one too cold,
and one just the right temperature.  So she eats the one that
is just right.  One of the beds is too hard, one too soft,
and one just the right degree of firmness.  So she lies down
on the one that is just right and falls asleep.  At this
point the three bears arrive home and $\ldots$ -- the
ending ranges from gruesome to happy depending on the version.

The Goldilocks principle states that, given a choice between
alternatives, one should choose the one that is `just right'
for the purpose at hand.  Applied to complex systems that
are adapted to a particular set of functions, the Goldilocks
principle states that there is a level of complexity that
is just right [12].  If the adapted system is too simple, then
it will fail to attain its various functions.  If it is too
complex, then it will be expensive and lack robustness.
When the complex system in question is one designed by
human beings -- a car, for example -- then the Goldilocks
principle is embodied by the widely applied theory of
axiomatic design [13-14].   In this theory, one looks
at the design matrix that relates input design parameters to
output functional parameters.  If one wishes to attain a
target set of functional parameters in a $D$-dimensional space,
then one needs to be able to vary at least $D$ design parameters
independently to attain that target.  More than $D$ design
parameters yields redundancy and unwanted complexity.

In the case of biological systems that have undergone
many generations of functional refinement via natural
selection, we expect the Goldilocks principle also to hold.
Biological systems evolve to the level of complexity required
to attain their necessary functions within their ecological
niche.  Having attained the requisite level of complexity,
they only add new features or increased complexity if by doing
so they can obtain a competitive advantage within that niche.

In this paper, we argue that energy transport in photosynthesis -- a
process that is, at bottom, quantum mechanical -- has attained the
level of quantum coherence and complexity that is just right for attaining high
efficiency (in some cases, almost 100\%) of transport of excitons
from the antenna to the reaction center.   In particular, the
interplay between coherent dynamics and environmental interaction
attains just the right level of quantum coherence to insure high
transport efficiency.   We show that the mechanism for this quantum
Goldilocks effect arises from the phenomenon of environmentally
assisted quantum transport (ENAQT), a mechanism that intrinsically
gives rise to efficient and robust transport over a range of the
various quantum timescales of the photosynthetic complexes [5-8].
Using a heuristic but quantitative argument, we show that the
transport efficiency is governed by a single parameter that is a
linear function of these timescales and the corresponding energy scales.  
Because of the linear
dependence of the efficiency parameter on the different timescales,
natural selection can arrive at optimal values for this parameter by
adjusting timescales in the proper direction until optimality is
attained -- that is, there are no sub-optimal local maxima to the
efficiency. Elsewhere, we show that this heuristic argument is
supported by detailed simulations of photosynthetic complexes [9-10]

\bigskip\noindent{\it Optimal timescales for quantum transport}

The fundamental mechanism that we propose governs the
efficiency of transport is environmentally assisted quantum
transport (ENAQT) [5-8].  ENAQT operates by
an interplay between coherence, decoherence, and disorder
in quantum systems.  Up to now, our
work on environmentally assisted quantum transport
has been based on detailed simulations of specific systems
such as FMO.  Here we provide a simple mechanism
for understanding quantitatively the interplay
between coherence, decoherence, and disorder.

Consider a quantum walker such as an exciton, moving through
an array of sites.  The array can be in one, two, or more dimensions;
it can possess a fractal dimension, or exhibit different
dimensions at different length scales.  It can be large or small.
The quantum walk can take place by local or by non-local interactions,
as in the dipolar interaction, as long as nearest neighbor
interactions are the strongest ones.

Let $J$ be the average strength of
coupling between neighboring sites, and assume that
the coupling between distant sites falls of relatively
rapidly (for example as in dipolar coupling).  Let $\ell$
be the transient localization length of the system: $\ell$ is the 
typical length of a path over which
the walker can propagate coherently before being
localized by destructive interference.  In a system that exhibits
Anderson localization [15], the transient localization length
is equal to the normal localization length.  However, as
discussed below, even systems that do not exhibit strict
Anderson localization still exhibit transient
localization.  The transient localization length
can be different in different parts of the system.  For
the analysis here, assume that $\ell > 1$, so that the 
walker can propagate at least one full step before being
localized.  The case of strong transient localization,
$\ell < 1$, will be discussed below.   The amount of
time it takes to propagate along the localization length
$\ell$ is $ \tau \approx \ell/2J$, (for
$J$ measured in Hertz).  

In the absence of environmental interactions, the system
propagates coherently over the localization time $\tau$
a distance equal to the localization length $\ell$,
and then becomes stuck due to destructive interference.
Decoherence can then `free up' the system and allow it to
propagate further.  Call the environmental decoherence rate
$d$.  The ENAQT mechanism suggests that there should
be an optimal rate of decoherence.

We estimate the optimal rate of decoherence by the following
simple argument. If the decoherence time is longer than the
localization time, $d < 1/\tau$, then the walker propagates
a length $\ell$, becomes stuck due to localization, and waits
around for a time $1/d$ to become unstuck.  At this point it
can propagate coherently for the localization length again, and the process
continues.  Such a process corresponds to a classical random
walk with step size $\ell$ and step time $1/d$, leading
to diffusion over time where $r(t) \approx \ell\sqrt{td}$.
In this regime, increasing the decoherence rate increases
the propagation rate.

Conversely, if the decoherence time is shorter than
the localization time, $d>\tau$, then
the walker propagates coherently
for a number of steps $2J/d < \ell$ before getting decohered.
By the same argument as before, it then
takes a classical random walk with step size $2J/d$ and step
time $1/d$, yielding diffusion over time
$r(t) \approx (2J/d)\sqrt{td} = (2J)\sqrt{t/d}$.
In this regime, decreasing the decoherence rate increases
the propagation rate.

The optimal rate of decoherence is seen to occur for $d\approx 1/\tau$.
That is, the decoherence time is equal to
the localization time -- no more, no less.  Setting
the decoherence time equal to the localization
time gives the interplay between coherence and decoherence
that is `just right.'  When $d\approx 1/\Delta t = 2J/\ell$,
the walker takes a classical random walk with step size $\ell$
and step time $1/d = \ell/2J = \Delta t$, yielding diffusion
over time as
$$r_{opt}(t) \approx \ell \sqrt{td} = \ell\sqrt{2tJ/ \ell} =
\sqrt{2tJ\ell}.\eqno(1)$$
Equation (1) gives the optimal rate of propagation
when the quantum Goldilocks principle is satisfied.

When the coupling between neighoring sites is 
less than the energy splitting between them, i.e.,
$J < \Delta \omega$, then coherent transport is strongly suppressed:
the walker has a relatively small amplitude of transiting to a
neighboring site.  A simple two-state model then shows
that the maximum probability of transition goes as
$J^2/ \Omega^2$, where $\Omega^2 = J^2 + \Delta\omega^2$,
and occurs at time $\Delta t = 1/2\Omega$.  As before,
the optimal decoherence rate goes as $d \approx 1/\Delta t$.
When $J<<\Delta \omega$, the optimal decoherence rate simply matches
the average energy disorder $\Delta \omega$
of the system.
At the optimal rate, takes a step with probability $J^2/ \Omega^2$
at time intervals, $\Delta t = 1/2\Omega$,
yielding random walk that spreads as
$J\sqrt{2t/\Omega}$.

Note the environmentally assisted quantum transport is an
intrinsically robust mechanism.  The optimum transport occurs when
the parameter $ \Lambda \equiv d \ell/2J \approx 1$, but optimality
extends over a considerable range of values of $\Lambda$. Our FMO
simulations show high efficiency of transport for $ 0.2 < \Lambda <
5$.

\bigskip\noindent{\it Transient localization}

As noted above, for our purposes it is not necessary that the system
exhibit strict Anderson localization [15].  Essentially any disordered
quantum system exhibits {\it transient} localization
over the time scale required for destructive interference
to build up.   Transient localization occurs even when
systems have long-range interactions, or exist in dimensions
where strict Anderson localization does not take place.
The transient localization length can be estimated as follows.
Localization arises from disorder in energy. 
Disorder in coupling constants and in the topology of the graph
also play a role.  For the moment consider the effect of
energy disorder, $\hbar\delta\omega$.
The transient localization length is a function
of $J/\Delta \omega$.  In particular, over the time scale
$1/2J$ required for the walker to propagate
from one site to the next, it acquires a random phase
$ \pm \pi \Delta \omega/J$.  After $n$ steps,
the walker acquires a phase $\pm \sqrt n \pi  \Delta \omega/J$.
Transient localization occurs when paths of propagation become sufficiently
long to acquire a random relative phase.  That is, transient localization
occurs when $\sqrt \ell \pi \Delta \omega/J \approx \pi$,
so that the transient localization length goes as
$\ell \approx (J/\Delta \omega)^2$ and the localization time
goes as $\tau \approx J/2\Delta^2$.  The optimal
decoherence rate is $d =  1/\tau= 2\Delta\omega^2/J$, and
the optimal transport rate goes as
$\sqrt{2tJ\ell} = (J/\Delta\omega)\sqrt{2tJ}$.
When disorder in coupling constants and in energy are included,
the transient localization length decreases and the estimate
$\ell \approx (J/\Delta \omega)^2$ is now an approximate upper
bound on the localization length.

Even when energies and couplings are identical, transient localization 
can also occur due to disorder in the network of sites through
which the walker is propagating, or due to quantum chaos [16-17].
As an example of transient localization, consider dipolar
interactions such as those that occur in photosynthesis.
Dipolar systems have long range interactions and will not be
localized in the long run.
Initially, a disordered quantum system exhibits coherent
propagation from the dominant,
short range part of the dipolar interaction.  Destructive
interference effectively freezes this propagation at a characteristic
time scale equal to the time needed for destructive interference
to build up.   Over much longer times, the long range part
of the interaction allows the walker to `leak out' of its
transiently localized state.  For excitons propagating
through photocomplexes, however, the time scale for the destruction
of transient localization by long range interactions is
typically longer than the exciton lifetime.  In the long
run, transient localization goes away.  But in the long run, the
excitons are all dead.

Transient localization also plays a role in systems that
are too small to exhibit strict Anderson localization.
For small systems, the transient localization length can be determined
by identifying the number of sites that participate in energy
eigenstates of the system.  The localization length can
differ from place to place within the system.  The average
energy splitting difference between the $\ell$ excitonic states in
a band delocalized
over $\ell$ sites is $ \Delta E \approx  2\pi\hbar J/\ell$.  Accordingly, our
quantum Goldilocks criterion $ d \ell/2J \approx 1$ that the
decoherence rate match the coherent propagation time is
equivalent to asking that the $ \pi\hbar d$ be equal
to the average energy splitting $\Delta E$ between exitonic states
in a band delocalized over $\ell$ sites.  

Phrased in terms of localized excitonic states, the criterion
$d\ell/2J \approx 1$ has simple explanation: it states that transport
is optimized when the local decoherence rate is approximately equal
to the local splitting in frequency between excitonic states in a 
band. That is, the optimal rate of decoherence is to `fuzz out' the local
energy levels until they begin to overlap with each other, and the
noise has sufficient spectral width to induce transitions between
excitonic states.  Note that this explanation does not rely on
relaxation.  Relaxation can also play an important role in
transport, particularly when the target state is at a lower energy
than the input states, as in FMO.  However, as shown in [5-10], the
interplay between coherence and pure decoherence is surprisingly
effective in arranging robust and efficient energy transport.

As noted above, localization lengths, energy splittings, and
decoherence rates can vary throughout a system.  When
there is such variation, optimal transport is attained
by matching
$$d \approx 2 J/\ell \approx \Delta E/\pi\hbar$$
locally throughout the system.

\bigskip\noindent{\it Convergence of time scales}

The theory of open quantum systems implies that the decoherence rate
at high temperatures -- which include the temperatures
physiologically relevant for photosynthesis -- 
$d = \alpha  \lambda kT/ \hbar^2 \gamma$, 
where $\lambda$ is the
reorganization energy which measures the strength of the interaction
between system and environment, $T$ is the temperature, $\gamma$
is the inverse bath correlation time, and the proportionality
constant $\alpha$ is $O(1)$ and depends upon the spectral density
of the environment, degree of non-Markovianity, etc.  
For example, an uncorrelated local
ohmic spectral density with frequency cutoff yields $\alpha = 2\pi$ [6].
This dependence of decoherence rate on fundamental parameters follows
in a straightforward way from second order perturbation theory in
the Markovian limit [6, 18-19].  However, as shown in [9-10, 19], it can
also capture the decoherence rate in non-perturbative, non-Markovian
scenarios, although with a different proportionality constant.  

The bath correlation length is also important:
if the fluctuations that induce decoherence in neighboring
sites are positively correlated, then the effective relative
decoherence rate is reduced.  Similarly, when the correlations
are negatively correlated, the effective relative decoherence rate
is increased.  Let $c$ be the normalized correlation coefficient
of the fluctuations, so that $c=1$ corresponds to perfect correlation
between neighboring sites, $c= -1$ corresponds to perfect anti-correlation,
and $c=0$ corresponds to no correlation.  Taking into account
the spatial correlations between sites yields a relative decoherence
rate between neighboring sites,
$d =  \alpha (1-c)\lambda kT/ \pi\gamma\hbar$.
Transport is optimized when
$d \approx 2J/\ell \approx \Delta E /\pi\hbar $,
where as above, $\Delta E$ is the average energy splitting of 
excitonic states in a band delocalized over $\ell$ sites.
The optimum occurs when the single parameter
$$\Lambda =   d \ell/2J = \alpha    (1-c) \lambda kT/ \hbar\gamma \Delta E 
\approx 1.  \eqno(2)$$

As noted, the analysis of optimality in quantum transport applies to
a wide variety of systems. The transport could take place over a
quasi-one-dimensional chain, as in a single LH1 or LH2 ring is
quasi-one-dimensional.  The couplings, disorder/energy splittings,
and decoherence rates need not be distributed uniformly throughout
the complex in general (with the exception of symmetric complexes
such as LH1 and LH2).  The above heuristic analysis should continue
to hold, however, as long as one looks at transport over individual
energy pathways within a complex.  For example, in LHCII, the
primary light-harvesting complex of green plants, the various energy
pathways within an LHCII monomer pass through sequences of
relatively localized and relatively delocalized states,
corresponding to more weakly coupled chromophores and to groups of
more strongly coupled chromophores, respectively. The analysis above
can then be applied to analyze the efficiency of transport along the
energy pathway step by step, as the exciton propagates incoherently
between weakly coupled chromophores, and then semi-coherently along
bands of delocalized states.  For each step, we expect transport to
be maximized for the decoherence rate approximately equal to the
disorder/energy difference associated with that step.

The complex and partially disordered quantum nature of
photocomplexes implies that a numerically accurate theoretical
picture of transport requires detailed simulations with the
appropriate open quantum system techniques, e.g., the
non-perturbative, non-Markovian master equation techniques mentioned
above. Applied to FMO, such simulations [9-10] are consistent with
the heuristic arguments given here.

\bigskip\noindent{\it Transport and the quantum Goldilocks effect}

As noted in [5-8] environmentally assisted quantum transport
intrinsically gives rise to high efficiencies over a relatively
wide range of parameters.  As long as the amount of time it
takes the exciton to diffuse throughout the photocomplex
is significantly less than the exciton lifetime, then
the efficiency will be close to one.  From equations (1-3), we see that
transport is optimized by matching the decoherence time with
the localization time, which is equivalent to
making the energy splitting within a localized band of
excitons comparable to the spreading of energies due to decoherence.
$X$-ray crystallography of photocomplexes, together with
quantum studies of excitonic states in chromophores, indicates
that chromophores are packed pretty much as closely as possible
within photocomplexes -- if they were packed more closely,
excitonic wave-function overlaps between chromophores would
lead to quenching and energy loss.  That is, naturally occurring
photosystems have evolved to a structure that effectively maximizes
the couplings between chromophores.
Detailed laser spectroscopy of photocomplexes suggests
the disorder/energy splitting in such complexes is indeed
comparable to the decoherence rate [2-4].

The degree of coherent propagation differs from
photocomplex to photocomplex.  The coherent propagation length
is largest in strongly coupled chromophore arrays
with a high degree of symmetry, such as $LH1$ and
$LH2$ rings, or the green sulphur bacteria chlorosome.  Larger
couplings and regular arrangements of chromophores
lead to excitonic states that are delocalized over a
larger number of sites, and the possibility for coherent propagation
of excitons between those sites. By contrast, the coherent propagation length
is smaller in more weakly coupled and
more diversely structured photocomplexes such as $FMO$ or $LHCII$,
where energy funnels also play a role.  Weaker coupling and
less regular structure lead to
more localized  states and to shorter coherent propagation
lengths.  In systems with significant energy differences between
input excitons and output states, relaxation can significantly
affect efficiency of transport, as well as decoherence.
Interestingly, however, our detailed simulations of FMO [9-10] show
that even when there are significant energy difference, the
interplay between coherence and decoherence can largely determine
the efficiency of transport.

As shown above, the efficiency of transport is governed
by the parameter $\Lambda = d\ell/2J$ for systems with a
localization length $ \ell$ considerably larger than one, 
and by $\Lambda = d/2\Omega$ for
highly localized systems, where $\Omega^2 = J^2 + \Delta \omega^2$.
FMO is a relatively highly localized system, with typical energy
splittings between chromophores larger than their couplings.
Engel {\it et al.} have measured the decoherence rate in FMO 
for quantum beating at
physiological temperatures [4], and have found values
$d=270 \pm 100 cm^{-1}$ for a transition with frequency 
$\Omega = 173 \pm 18 cm^{-1}$, yielding $\Lambda
\approx 0.8 \pm 0.3$, consistent with the value 1.

We would like to estimate $\Lambda$ for systems where the transient
localization length $\ell$ is greater than one, for
which $\Lambda = \tau d = \ell d/2J$.  In the 850 nanometer ring 
of LH2, for example, excitons are delocalized over multiple
sites in the ring.  Str\"umpfer and Schulten [19] calculate
the time it takes an exciton to spread from a single site
to half the ring (nine sites) as $\approx 150$ femtoseconds,
suggesting a localization time of $\approx 100$ femtoseconds
if the localization length is somewhat less
than half the ring.  That is, the strong coupling between 
chromophores in LH2 make for longer transient localization
lengths and shorter localization times than in FMO.
We can now use the generic form of 
the decoherence rate at physiological temperatures $T$
$d = \alpha (1-c)\lambda kT/\hbar^2\gamma$ to estimate
$\Lambda$ in LH2.
Str\"umpfer and Schulten [20] report the following
values for FMO:
$\lambda=35 {\rm cm}^{-1}, \gamma=50 {\rm cm}^{-1}$.
For the 850 nm ring of LH2, they report
$\lambda= 200{\rm cm}^{-1}, \gamma=83 {\rm cm}^{-1}$,
suggesting a decoherence rate several times faster for LH2
than for FMO (assuming that the proportionality constant
$\alpha$ takes on similar values for the two distinct systems).
But as just noted, they also estimate the coherent propagation
time to be several times shorter for LH2 than for FMO,
once again suggesting $\Lambda \approx 1$. 

The green sulphur bacteria chlorosome consists of regular
cylindrical arrays of tightly coupled chromophores, which
might exhibit coherent propagation of excitons over significant
scales.  Simulations of the interplay of coherence and decoherence
in cylindrical and spiral arrays of chromophores in the presence
of disorder show clear evidence for the destruction of localization
via the addition of decoherence [21], with optimal transport for decoherence
rates approximately equal to the inverse localization time.

The reorganization energy, correlation time of environment,
and disorder/energy splitting, are all parameters that have
the potential to be tuned by natural selection.
Decoherence rates differ throughout the photocomplex: our
model predicts that these rates -- if optimized by natural
selection -- should correspond to
to the exciton splitting in localized bands.
The simple form of $\Lambda$ as the ratio of
fundamental quantum parameters shows that optimum transport
can be attained simply be evolving each of the parameters in
the right direction until $\Lambda = 1$ is reached -- there
are no false maxima at relatively low efficiency.

\bigskip\noindent{\it Discussion}

This paper discussed the divergence and convergence of time and
energy scales in quantum transport processes.  The convergence
of time scales can be thought of as an example of the
quantum Goldilocks effect, which states that there is a
`just right' level of quantum complexity and coherence for functional
complex quantum systems such as photosynthetic complexes.
When the time scales for different quantum processes converge,
these processes can help each other out to give more efficient transport.

We then analyzed a potential mechanism of attaining robust,
efficient quantum transport. Environmentally assisted quantum
transport (ENAQT) uses the interplay between coherence and
decoherence to attain efficiency and robustness [5-10].   We gave a
heuristic but quantitative argument that showed that the efficiency
of transport can be increased (a) by increasing the coupling between
sites, (b) by minimizing disorder and energy splitting, and (c) by
setting the decoherence rate equal to the disorder/energy splitting.
This heuristic argument complements the detailed simulations
presented in [9-10,18]. We showed that transport efficiency at high
temperatures is a function of the single parameter $\Lambda = d
\ell/2J \approx \alpha  (1-c) \pi \lambda kT/\hbar \gamma  \Delta E$, 
where $d$ is
the decoherence rate, $\ell$ is the transient localization length,
$J$ is the local coupling, $c$ is the correlation coefficient for
noise between neighboring states, $\lambda$ is the reorganization
energy, $T$ is the temperature, $\gamma$ is the inverse correlation
time of the environment, and $\Delta E$ is the energy splitting
within a localized band of states, and $\alpha$ is a constant
of order one. High efficiency is attained for
$\Lambda \approx 1$. This simple set of rules for increasing quantum
transport efficiency together with the simple form for the governing
parameter $\Lambda$, provides a straightforward path for natural
selection to implement the quantum Goldilocks effect -- there is
only one broad optimum for the ratios of values, and changing any
one of the timescales in the system to make $\Lambda$ closer to one
will typically increase efficiency. While this paper focused on
excitonic transport in photosynthetic complexes, the heuristic,
quantitative analysis given can be applied to any quantum transport
process, in relatively complex quantum systems that exhibit a
combination of order and disorder.  
The quantum Goldilocks effect
can be used to maximize efficiency and minimize energy
dissipation in any process in which a quantum system
progresses through a sequence of quantum-mechanical states
in order to accomplish some function.  For example, conformational
changes in proteins or in optically active molecules such as
retinal can be accelerated by the proper balance of coherence
and decoherence.  The quantum Goldilocks effect is not confined
to systems that have been optimized by natural selection: the principles
for maximizing transport efficiency described here for evolved
systems could also fruitfully be applied to the design of artificial
energy harvesting systems.

\vfill
\noindent{\it Acknowledgements:}  We dedicate this work to our
colleague Bob Silbey.   We thank Silbey, Jianshu Cao, Greg Engel,
Greg Scholes, and Jake Taylor for helpful discussions.
This work was supported by
ENI under the MIT Energy Initiative, DARPA under the QuBE
program, Intel, NEC, Lockheed Martin, and NSF.

\vfil\eject

\bigskip\noindent{\it References:}

\bigskip\noindent [1] R.E. Blankenship,  {\it Molecular Mechanisms of
Photosynthesis,} London: Blackwell Science (2002).

\bigskip\noindent [2]
G.S. Engel, T.R. Calhoun, E.L. Read, T.K. Ahn, T. Mancal, 
 Y.C. Cheng, R.E. Blankenship, G.R. Fleming, 
{\it Nature} {\bf 446}, 782-786 (2007).

\bigskip\noindent [3]
E. Collini, C.Y. Wong, K.E. Wilk, P.M. Curmi, P. Brumer, G.D.
Scholes, 
{\it Nature} {\bf 463}, 644-647 (2010).

\bigskip\noindent [4]
G. Panitchayangkoon, D. Hayes, K.A. Fransted, J.R. Caram, 
E. Harel, J. Wen, R.E. Blankenship, G.S. Engel, 
{\it Proc. Nat.
Acad. Sci.} {\bf 107}, 12766-12770 (2010).

\bigskip\noindent [5]
M. Mohseni, P. Rebentrost, S. Lloyd, A. Aspuru-Guzik, 
{\it J.  Chem. Phys.} {\bf 129}, 174106 (2008).

\bigskip\noindent [6]
P. Rebentrost, M. Mohseni, I. Kassal, S. Lloyd, A. Aspuru-Guzik,
Environment-assisted quantum transport. {\it New J. Phys.} {\bf 11}, 033003
(2009).

\bigskip\noindent [7] P. Rebentrost, M. Mohseni, A. Aspuru-Guzik, 
{J. Phys. Chem. B} {\bf 113}, 9942-9947 (2009).

\bigskip\noindent [8]
M.G. Plenio, S.F. Huelga, 
{\it New J. Phys.} {\bf  10}, 113019 (2008).

\bigskip\noindent [9] A. Shabani, M. Mohseni, H. Rabitz, S. Lloyd, 
(I) Efficient simulation of excitonic dynamics in the non-perturbative
and non-Markovian regimes,' arXiv:1103.3823; submitted to {J. Phys.
Chem.} (2011).

\bigskip\noindent [10] M. Mohseni, A. Shabani, S. Lloyd, H. Rabitz,
`Optimal and robust energy transport in light-harvesting complexes:
(II) A quantum interplay of multichromophoric geometries and
environmental interactions,'  arXiv:1104.4812;  submitted to {J. Phys. 
Chem.} (2011).

\bigskip\noindent [11] J. Wu, F. Liu, Y. Shen,  J. Cao, 
R.J. Silbey, {\it New J. Phys.} {\bf 12} 105012 (2010).

\bigskip\noindent [12]
S. Lloyd, lecture at Santa Fe Institute, 1990, as reported by
M. Gell-Mann, {\it The Quark and the Jaguar,} A. Knopf, New York
(1994).

\bigskip\noindent [13]
N.P. Suh, {\it The Principles of Design}, Oxford University Press,
Oxford (1990).

\bigskip\noindent [14]
N.P. Suh, {\it Axiomatic Design: Advances and Applications,}
Oxford University Press, Oxford (2001).

\bigskip\noindent [15] P.W. Anderson, {\it Phys. Rev.} {\bf 109}, 1492–1505
(1958).

\bigskip\noindent [16] G. Casati, I. Guarneri, P.L. Shepelyansky,
{\it IEEE J. Quant. Elect.} {\bf 24}, 1420-1444 (1988).

\bigskip\noindent [17] G. Casati, V. Chirikov, {\it Quantum 
Chaos: Between order and disorder}, Cambridge University Press,
Cambridge, 1995.

\bigskip\noindent [18]
H.-P. Breuer, F. Petruccione, {\it The theory of open quantum
systems}, Oxford University Press, Oxford (2007).

\bigskip\noindent [19]
A. Ishizaki, G.R. Fleming, 
 {\it J. Chem. Phys.} {\bf 130}, 234111 (2009).

\bigskip\noindent [20] J. Str\"umpfer, K. Schulten, 
{\it J. Chem. Phys.} {\bf 131}, 225101 (2009).

\bigskip\noindent [21] D.F. Abasto, M. Mohseni, S. Lloyd, 
P. Zanardi, Excitonic diffusion length in complex quantum 
systems: The effects of disorder and environmental fluctuations 
on symmetry-enhanced supertransfer, to appear in {\it Phil.
Trans. Roy. Soc.}, arXiv:1105.418 (2011).

\vfill\eject\end